\documentclass[referee]{elsarticle}

\usepackage{graphicx}
\usepackage{bm}
\usepackage{amsmath,amssymb}

\journal{Physics Letters A}

\begin{document}
\begin{frontmatter}
\title{%
Wave turbulent statistics
in non-weak wave turbulence
}

\author{Naoto Yokoyama\corref{cor}}
\address{
Department of Aeronautics and Astronautics,
Kyoto University,
Kyoto 606-8501 JAPAN
}
\cortext[cor]{Tel: +81-774-38-3961, Fax: +81-774-38-3962}
\ead{yokoyama@kuaero.kyoto-u.ac.jp}

\date{\today}

\begin{abstract}
In wave turbulence,
it has been believed that
statistical properties are well described by the weak turbulence theory,
in which nonlinear interactions among wavenumbers are assumed to be small.
In the weak turbulence theory,
separation of linear and nonlinear time scales derived from the weak nonlinearity
is also assumed.
However, the separation of the time scales is often violated even in weak turbulent systems
where the nonlinear interactions are actually weak.
To get rid of this inconsistency,
closed equations are derived without assuming the separation of the time scales
in accordance with Direct-Interaction Approximation (DIA),
which has been successfully applied to Navier--Stokes turbulence.
The kinetic equation of the weak turbulence theory is recovered from the DIA equations
if the weak nonlinearity is assumed as an additional assumption.
It suggests that the DIA equations is a natural extension
of the conventional kinetic equation
to {\em not-necessarily-weak\/} wave turbulence.
\end{abstract}


\end{frontmatter}

\section{Introduction}\label{sec:introduction}

Energy in wave fields such as ocean surface waves~\cite{zak1968}, ocean internal waves~\cite{iwthLPTN}, and elastic waves on thin metal plates~\cite{during2006weak}
is transferred among wavenumbers owing to nonlinear interactions.
The large degree-of-freedom wave fields
are called wave turbulence.
The weak turbulence theory,
in which the nonlinear interactions are assumed to be small,
is partially successful in wave turbulent statistics.
Therefore,
``the {\em wave\/} turbulence'' and ``the {\em weak\/} turbulence'' are often regarded as synonyms.

However,
the separation of the linear and nonlinear time scales,
which is assumed in the weak turbulence theory,
is often violated as pointed out based on the weak turbulence theory~\cite{newell01,Connaughton200386,During20091524}.
Especially in anisotropic wave turbulence such as
ocean internal waves~\cite{iwthLPTN} and Alfv\'en waves~\cite{nazarenko2001nlm},
the separation of the time scales is almost always violated.
The violations are often observed
also in both observations and direct numerical simulations
in ocean surface waves~\cite{naoto_jfm} and ocean internal waves~\cite{numericsLY}.
In the direct numerical simulations,
the violations appear to be caused by fast non-resonant interactions.
Therefore, the wave turbulence theory applicable to the non-weak wave turbulence is required
for statistics of the not-necessarily-weak wave turbulence.

In Navier--Stokes turbulence which has strong nonlinear interactions among wavenumbers,
direct-interaction approximation (DIA)~\cite{kraichnan5hsi}
is considered to successfully describe
turbulent statistics of Navier--Stokes turbulence.
In this Letter,
DIA is applied to a three-wave turbulent system,
and a wave turbulence theory that accepts short-time and strong nonlinear interactions
is constructed.

\section{Weak Turbulence Theory}
\label{sec:WT}

Wave turbulent systems with nonlinear interactions among three wavenumbers
is generally governed by a canonical equation 
\begin{align}
 \mathrm{i} \frac{\partial a(\bm{p})}{\partial t} &=
\frac{\delta \mathcal{H}}{\delta a^{\ast}(\bm{p})}
\label{eq:generalcanonical}
\end{align}
for a complex amplitude $a(\bm{p})$
with Hamiltonian
\begin{align*}
 \mathcal{H} = & 
\ 
(2\pi)^{-d} \! \int \! \mathrm{d}\bm{p} \ \omega |a(\bm{p})|^2
\\
&
\!\!\!\!
+ (2\pi)^{-2d} \! \int \! \mathrm{d}\bm{p} \ \mathrm{d}\bm{p}_1 \mathrm{d}\bm{p}_2
\left(
\mathcal{T}^{\bm{p}}_{\bm{p}_1\bm{p}_2} a^{\ast}(\bm{p}) a(\bm{p}_1) a(\bm{p}_2)
+ \mathrm{c.c.}
\right),
\end{align*}
where $\bm{p}$ is a $d$-dimensional wavenumber vector,
and
$\omega(\bm{p})$ is a frequency of wavenumber $\bm{p}$ given by a linear dispersion relation.
A matrix element $\mathcal{T}^{\bm{p}}_{\bm{p}_1\bm{p}_2}$ gives strength of three-wave nonlinear interactions among $\bm{p}$, $\bm{p}_1$ and $\bm{p}_2$.
A functional derivative with respect to $a$ is expressed by $\delta / \delta a$.
Moreover,
$a^{\ast}$ denotes the complex conjugate of $a$,
and $\mathrm{c.c.}$ also denotes the complex conjugate of the previous term.
A $d$-dimensional periodic wave field in a volume $(2\pi)^d$
is considered for simplicity.
Since the wave media are assumed to have spacial inversion symmetry,
$\omega(\bm{p}) = \omega(-\bm{p})$
and
$\mathcal{T}^{\bm{p}}_{\bm{p}_1\bm{p}_2} = \mathcal{T}^{-\bm{p}}_{-\bm{p}_1 -\bm{p}_2}$.

The canonical equation~(\ref{eq:generalcanonical}) is rewritten as
\begin{align}
& \frac{\partial a(\bm{p})}{\partial t} =
- \mathrm{i} \omega(\bm{p}) a(\bm{p})
\nonumber\\
&
 -\mathrm{i}
 \!\!\!\!\!\! \sum_{\bm{p} = \bm{p}_1 + \bm{p}_2} \!\!\!\!\!\!
 \mathcal{T}^{\bm{p}}_{\bm{p}_1 \bm{p}_2} a(\bm{p}_1) a(\bm{p}_2)
 - 2\mathrm{i} \!\!\!\!\!\! \sum_{\bm{p} = - \bm{p}_1 + \bm{p}_2} \!\!\!\!\!\!
{\mathcal{T}^{\bm{p}_2}_{\bm{p} \bm{p}_1}}^{\ast} a^{\ast}(\bm{p}_1) a(\bm{p}_2)
.
\label{eq:adot}
\end{align}
If the nonlinear terms in Eq.~(\ref{eq:adot}) is neglected,
it can easily be integrated as
\begin{align*}
 a(\bm{p}) = A(\bm{p}) \exp(-\mathrm{i}\omega(\bm{p}) t +\mathrm{i}\theta(\bm{p}))
.
\end{align*}
Therefore,
$a(\bm{p})$ is called complex amplitude
and represents the behaviour of a mode $\bm{p}$ in the phase space.

The turbulent statistics governed by Eq.~(\ref{eq:adot})
is conventionally given
according to the weak turbulence theory~\cite{zak_book,janssen_freak}.
Equation~(\ref{eq:adot}) is multiplied by $a^{\ast}(\bm{p}_4)$
and
added complex conjugate with $\bm{p}$ and $\bm{p}_4$ interchanged,
and the ensemble averaging ($\overline{\cdots}$) is taken:
\begin{align}
 \frac{\partial n(\bm{p})}{\partial t} \delta_{\bm{p} \bm{p}_4}
 = & -\mathrm{i} \!\!\!\! \sum_{\bm{p} = \bm{p}_1 + \bm{p}_2} \!\!\!\!
 \mathcal{T}^{\bm{p}}_{\bm{p}_1 \bm{p}_2} \overline{a^{\ast}(\bm{p}_4) a(\bm{p}_1) a(\bm{p}_2)}
\nonumber\\
&
 - 2\mathrm{i} \!\!\!\! \sum_{\bm{p} = - \bm{p}_1 + \bm{p}_2} \!\!\!\! 
{\mathcal{T}^{\bm{p}_2}_{\bm{p} \bm{p}_1}}^{\ast} \overline{a^{\ast}(\bm{p}_4) a^{\ast}(\bm{p}_1) a(\bm{p}_2)}
\nonumber  \\
  &
  + \mathrm{c.c.} (\bm{p} \leftrightarrow \bm{p}_4)
,
\label{eq:2to3}
 \end{align}
where
$\mathrm{c.c.}(\bm{p} \leftrightarrow \bm{p}_4)$
denotes taking complex conjugates
with $\bm{p}$ and $\bm{p}_4$ interchanged.
Kronecker's delta is denoted as $\delta_{\bm{p}_i \bm{p}_j}$.
The wave action $n(\bm{p})$,
which is energy of wavenumber $\bm{p}$ divided by $\omega$,
is defined by $\overline{a(\bm{p}_i) a^{\ast}(\bm{p}_j) } = n(\bm{p}_i) \delta_{\bm{p}_i \bm{p}_j}$.

By applying the random phase approximation in the weak turbulence theory
in the most primitive sense
$\overline{ a(\bm{p}_i) a^{\ast}(\bm{p}_j) a^{\ast}(\bm{p}_k) } = 0$
to the third-order correlation,
the nonlinear term vanishes
and
\begin{align*}
 \frac{\partial n(\bm{p})}{\partial t} = 0
.
\end{align*}
Then,
the non-zero third-order correlation is obtained
by considering the time evolution of the third-order correlation.
The time evolution of the third-order correlation is written as
\begin{align}
 \left( \frac{\partial}{\partial t} +\mathrm{i} \Delta \omega^{\bm{p}}_{\bm{p}_1\bm{p}_2} \right) \overline{ a a_1^{\ast} a_2^{\ast} }
 &= -\mathrm{i}
 \left(
 \sum_{\bm{p} = \bm{p}_3 + \bm{p}_4} 
\!\!\!\!\!\!
 \mathcal{T}^{\bm{p}}_{\bm{p}_3\bm{p}_4} \overline{ a_1^{\ast} a_2^{\ast} a_3 a_4 }
 + 2
\!\!\!\!\!\!
 \sum_{\bm{p} + \bm{p}_3 = \bm{p}_4}
\!\!\!\!\!\!
 \mathcal{T}^{\bm{p}_4 \ast}_{\bm{p}\bm{p}_3} \overline{ a_1^{\ast} a_2^{\ast} a_3^{\ast} a_4 }
 \right)
\nonumber \\
 &
\qquad
 + \mathrm{c.c.} (\bm{p} \leftrightarrow \bm{p}_1)
 + \mathrm{c.c.} (\bm{p} \leftrightarrow \bm{p}_2)
\nonumber \\
 &= -2\mathrm{i} \mathcal{T}^{\bm{p}}_{\bm{p}_1\bm{p}_2}
 \left(n_1 n_2 - n (n_1 + n_2)\right)
.
\label{eq:c3}
\end{align}
The random phase approximation is applied to the fourth-order correlation:
\begin{align*}
& \overline{ a_{\bm{p}_i} a_{\bm{p}_j} a^{\ast}_{\bm{p}_k} a^{\ast}_{\bm{p}_l} }
 =
 \overline{a_{\bm{p}_i} a^{\ast}_{\bm{p}_k}} \, \overline{a_{\bm{p}_j} a^{\ast}_{\bm{p}_l} }
 +
 \overline{a_{\bm{p}_i} a^{\ast}_{\bm{p}_l}} \, \overline{a_{\bm{p}_j} a^{\ast}_{\bm{p}_k}  }
 \nonumber\\
& = n(\bm{p}_i) n(\bm{p}_j)
 \left(
 \delta_{\bm{p}_i \bm{p}_k} \delta_{\bm{p}_j \bm{p}_l}
 +
 \delta_{\bm{p}_i \bm{p}_l} \delta_{\bm{p}_j \bm{p}_k}
 \right)
.
\end{align*}
The frequency difference among the three wavenumbers is denoted by $\Delta \omega^{\bm{p}}_{\bm{p}_1\bm{p}_2} = \omega(\bm{p}) - \omega(\bm{p}_1) - \omega(\bm{p}_2)$.

The time variation of wave action $n$ appearing in the right-hand side of Eq.~(\ref{eq:c3})
is possibly negligible
by comparing with $1/\Delta \omega^{\bm{p}}_{\bm{p}_1\bm{p}_2}$
if the nonlinearity is weak.
Here,
the linear timescales are assumed to be much faster than the nonlinear time scales.
The separation of the time scales is nontrivial and is often violated.

When the separation of the time scales are valid,
Eq.~(\ref{eq:c3}) can be integrated from $t_0$ to $t_0+\tau$,
under the initial condition $\overline{a a_1^{\ast} a_2^{\ast} }(t_0)=0$.
The third-order correlation is obtained as
\begin{align*}
 \overline{ a a_1^{\ast} a_2^{\ast} } = \frac{-2\mathrm{i} \mathcal{T}^{\bm{p}}_{\bm{p}_1 \bm{p}_2}
 \left(n_1 n_2 - n (n_1 + n_2)\right) \left(\exp(-\mathrm{i} \Delta \omega^{\bm{p}}_{\bm{p}_1\bm{p}_2} \tau)-1\right)}{-\mathrm{i}\Delta \omega^{\bm{p}}_{\bm{p}_1\bm{p}_2}}
.
\end{align*}
By employing
\begin{align*}
 \frac{\mathrm{i} (\exp(-\mathrm{i}\Delta \omega \tau)-1)}{-\mathrm{i}\Delta \omega}
 =
 \mathrm{P.V.}\left(\frac{1}{\Delta \omega}\right) +\mathrm{i}\pi \delta(\Delta \omega)
 &
\qquad
 \text{as } \tau \to \infty
,
\end{align*}
Eq.~(\ref{eq:c3}) finally results in the kinetic equation:
 \begin{align}
  \frac{\partial n}{\partial t}
  = & \ 4 \pi
\left(
\sum_{\bm{p} = \bm{p}_1 + \bm{p}_2}
\!\!\!\!
    |\mathcal{T}^{\bm{p}}_{\bm{p}_1\bm{p}_2}|^2 \left(n_1 n_2 - n (n_1 + n_2)\right)
\delta(\Delta \omega^{\bm{p}}_{\bm{p}_1\bm{p}_2})
  \right.
  \nonumber\\
  &
  - 
\!\!\!\!
\sum_{\bm{p}_1 = \bm{p}_2 + \bm{p}}
\!\!\!\!
|\mathcal{T}^{\bm{p}_1}_{\bm{p}_2\bm{p}}|^2 \left(n_2 n - n_1 (n_2 + n)\right)
\delta(\Delta \omega^{\bm{p}_1}_{\bm{p}_2\bm{p}})
  \nonumber\\
  &
  \left.
  - 
\!\!\!\!
\sum_{\bm{p}_2 = \bm{p} + \bm{p}_1}
\!\!\!\!
|\mathcal{T}^{\bm{p}_2}_{\bm{p}\bm{p}_1}|^2 \left(n n_1 - n_2 (n + n_1)\right)
\delta(\Delta \omega^{\bm{p}_2}_{\bm{p}\bm{p}_1})
  \right)
.
  \label{eq:kineticeq}
 \end{align}
The kinetic equation indicates that
the energy is transferred
among wavenumbers which satisfy the resonant conditions:
\begin{align*}
 \begin{cases}
  \bm{p} = \bm{p}_1 + \bm{p}_2
  \\
  \omega = \omega_1 + \omega_2
 \end{cases}
\!\!\!\!\!\!\!
,
\ \ \ \ 
 \begin{cases}
  \bm{p}_1 = \bm{p}_2 + \bm{p}
	    \\
  \omega_1 = \omega_2 + \omega
 \end{cases}
\!\!\!\!\!\!
,
\ \ \ \ 
 \begin{cases}
  \bm{p}_2 = \bm{p} + \bm{p}_1
  \\
  \omega_2 = \omega + \omega_1
 \end{cases}
\!\!\!\!\!\!\!
.
\end{align*}

\section{Direct-Interaction Approximation for Wave Turbulence}
\label{sec:DIA}

As written in \S\ref{sec:WT},
the weak turbulence theory assumes,
derived from the weak nonlinearity,
that
the time scales of the nonlinear energy transfer is much larger than
the linear time scales.
However,
in almost all the anisotropic wave turbulence,
the kinetic equation~(\ref{eq:kineticeq}) itself gives
short-time strong nonlinear interactions,
and the separation of the linear and nonlinear time scales are often violated.

In this section,
the direct-interaction approximation (DIA)~\cite{goto_kida_weakness},
in which the separation of the time scales is not assumed
but the largeness of the degrees of freedom is assumed,
is applied to the wave turbulence.
Note that the largeness of the degrees of freedom is also assumed implicitly in the weak turbulence theory.
Instead of complex amplitude $a$,
variables $b_1$ and $b_2$
defined by
\begin{align*}
 b_1(\bm{p}) = \frac{a(\bm{p}) + a^{\ast}(-\bm{p})}{2},
\quad
 b_2(\bm{p}) = \frac{a(\bm{p}) - a^{\ast}(-\bm{p})}{2\mathrm{i}}
,
\end{align*}
are employed.
For example,
the variables are defined as
\begin{align*}
 b_1(\bm{p}) = \sqrt{\frac{\sigma |\bm{p}|}{2\rho}} \eta(\bm{p})
,
\quad
 b_2(\bm{p}) = \sqrt{\frac{\rho}{2\sigma|\bm{p}|}} \phi(\bm{p})
,
\end{align*}
where $\eta(\bm{p})$ and $\phi(\bm{p})$ are
the Fourier transform of the surface elevation and that of the velocity potential
in deep capillary waves~\cite{zak_book},
and they are
\begin{align*}
b_1(\bm{p}) = \frac{\sqrt{\omega} N_0}{\sqrt{2g} |\bm{k}|} \Pi(\bm{p})
,
\quad
b_2(\bm{p}) = -\frac{\sqrt{g} |\bm{k}|}{\sqrt{2\omega}N_0} \phi(\bm{p})
,
\end{align*}
where $\Pi(\bm{p})$ and $\phi(\bm{p})$ are
the Fourier transform of the stratification thickness
and that of the velocity potential in ocean internal waves~\cite{lvov-2001-87}.
Since $b_i(\bm{p})$ is the Fourier transform of real functions,
$b_i(\bm{p}) = b_i^{\ast}(-\bm{p})$.

The governing equation~(\ref{eq:adot}) is rewritten as
\begin{align}
 \frac{\partial b_i(\bm{p})}{\partial t} &= \mathcal{L}_{ij}(\bm{p}) b_j(\bm{p})
 +
 \!\!\!\!\!\!\!\!\!\!
 \sum_{\bm{p} + \bm{p}_1 + \bm{p}_2 = \bm{0}}
 \!\!\!\!\!\!\!\!\!
 \mathcal{N}_{ijk}(\bm{p},\bm{p}_1,\bm{p}_2)
 b_j(-\bm{p}_1) b_k(-\bm{p}_2)
.
\label{eq:governingeqb}
\end{align}
The Einstein summation convention is employed.
The linear and nonlinear matrix elements are
\begin{align*}
\mathcal{L}_{ij}(\bm{p}) &= (j-i) \omega(\bm{p}),
\nonumber\\
\mathcal{N}_{ijk}(\bm{p},\bm{p}_1,\bm{p}_2) &= \frac{\mathrm{i}^{i+j+k}}{2} 
\left(
(-1)^{i+1} \mathcal{T}_{-\bm{p}_1 -\bm{p}_2}^{\bm{p}}
+ (-1)^j \mathcal{T}_{-\bm{p} -\bm{p}_2}^{\bm{p}_1}
+ (-1)^k \mathcal{T}_{-\bm{p} -\bm{p}_1}^{\bm{p}_2}
\right)
\nonumber\\
&
\qquad
+ \left(
\mathrm{c.c.}(\bm{p}, \bm{p}_1, \bm{p}_2) \to (-\bm{p}, -\bm{p}_1, -\bm{p}_2)
\right)
,
\end{align*}
where
$\mathrm{c.c.}(\bm{p}, \bm{p}_1, \bm{p}_2) \to (-\bm{p}, -\bm{p}_1, -\bm{p}_2)$
denotes taking complex conjugates
and replacing $(\bm{p}$, $\bm{p}_1$, $\bm{p}_2)$ by
$(-\bm{p}$, $-\bm{p}_1$, $-\bm{p}_2)$ simultaneously.

When a perturbation is added
to the $j$th component of a wavenumber $\bm{p}^{\prime}$
at a time $t^{\prime}$,
which is $b_j(\bm{p}^{\prime}, t^{\prime})$,
the $i$th component of another wavenumber $\bm{p}$ at a later time $t$,
which is $b_i(\bm{p}, t)$,
responds to the perturbation.
The response function is defined as
\begin{align*}
 G_{ij}(\bm{p}, t | \bm{p}^{\prime}, t^{\prime}) = \frac{\delta b_i(\bm{p}, t)}{\delta b_j(\bm{p}^{\prime}, t^{\prime})}
.
\end{align*}
From Eq.~(\ref{eq:governingeqb}),
the governing equation of the response function is given by
\begin{align*}
& \frac{\partial G_{in}(\bm{p}, t | \bm{p}^{\prime}, t^{\prime})}{\partial t} = \mathcal{L}_{ij}(\bm{p}) G_{jn}(\bm{p}, t | \bm{p}^{\prime}, t^{\prime})
\nonumber\\
&
 + 2 
 \!\!\!\!\!\!\!\!
 \sum_{\bm{p} + \bm{p}_1 + \bm{p}_2 = \bm{0}}
 \!\!\!\!\!\!\!\!
 \mathcal{N}_{ijk}(\bm{p},\bm{p}_1,\bm{p}_2)
 b_j(-\bm{p}_1) G_{kn}(-\bm{p}_2, t | \bm{p}^{\prime}, t^{\prime})
.
\end{align*}
The initial condition of the response function is given as
\begin{align*}
G_{ij}(\bm{p}, t^{\prime} | \bm{p}^{\prime}, t^{\prime})
 = \delta_{ij} \delta(\bm{p} - \bm{p}^{\prime})
.
\end{align*}

A perturbation that removes
a triad interaction among $\bm{p}_0$, $\bm{q}_0$ and $\bm{r}_0$
that satisfies $\bm{p}_0 + \bm{q}_0 + \bm{r}_0 = \bm{0}$
is added to a wave field at a time $t^{\prime}$.
When the degrees of freedom is large enough, the effect of the perturbation is small.
The variable $b_i$ is resolved into 
no direct-interaction field (NDI)
where the direct interaction
among $\bm{p}_0$, $\bm{q}_0$ and $\bm{r}_0$ is removed
and direct-interaction field (DI).
Then, $b_i = b_i^{(0)} + b_i^{(1)}$
and
$|b_i^{(0)}| \gg |b_i^{(1)}|$.
Similarly,
the response function is also resolved into 
$G_{ij}^{(0)}$ in NDI
and $G_{ij}^{(1)}$ in DI:
$G_{ij} = G_{ij}^{(0)} + G_{ij}^{(1)}$
and
$|G_{ij}^{(0)}| \gg |G_{ij}^{(1)}|$.

The equation of $b^{(0)}$ is given as
\begin{align*}
& \frac{\partial b_i^{(0)}(\bm{p})}{\partial t} = \mathcal{L}_{ij}(\bm{p}) b_j^{(0)}(\bm{p})
 + 
 \!\!\!\!\!\!\!\! \!\!\!\!\!\!\!\!
\sum_{\substack{\bm{p} + \bm{p}_1 + \bm{p}_2 = \bm{0} \\
 \{\bm{p},\bm{p}_1,\bm{p}_2\} \neq \{\bm{p}_0,\bm{q}_0,\bm{r}_0\} }}
 \!\!\!\!\!\!\!\! \!\!\!\!\!\!\!\!
 \mathcal{N}_{ijk}(\bm{p},\bm{p}_1,\bm{p}_2)
 b_j^{(0)}(-\bm{p}_1) b_k^{(0)}(-\bm{p}_2)
,
\end{align*}
and
that of $b^{(1)}$ is
\begin{align*}
& \frac{\partial b_i^{(1)}(\bm{p})}{\partial t} = \mathcal{L}_{ij}(\bm{p}) b_j^{(1)}(\bm{p})
\nonumber\\
&
 + 2
 \!\!\!\!\!\!\!\! \!\!\!\!\!\!\!\!
 \sum_{\substack{\bm{p} + \bm{p}_1 + \bm{p}_2 = \bm{0} \\
 \{\bm{p},\bm{p}_1,\bm{p}_2\} \neq \{\bm{p}_0,\bm{q}_0,\bm{r}_0\} }}
 \!\!\!\!\!\!\!\! \!\!\!\!\!\!\!\!
 \mathcal{N}_{ijk}(\bm{p},\bm{p}_1,\bm{p}_2)
 b_j^{(0)}(-\bm{p}_1) b_k^{(1)}(-\bm{p}_2)
\nonumber\\
&+
 \mathcal{N}_{ijk}(\bm{p}_0,\bm{q}_0,\bm{r}_0)
\left(
\delta_{\bm{p} \bm{p}_0}
 b_j^{(0)}(-\bm{q}_0) b_k^{(0)}(-\bm{r}_0)
+
\delta_{\bm{p} \bm{p}_0}
 b_j^{(0)}(\bm{q}_0) b_k^{(0)}(\bm{r}_0)
\right)
\nonumber\\
&
+
\{\bm{p}_0, \bm{q}_0, \bm{r}_0\}
.
\end{align*}
Here,
$\{\bm{p}_0, \bm{q}_0, \bm{r}_0\}$ denotes cyclic permutations
from $(\bm{p}_0, \bm{q}_0, \bm{r}_0)$.
The governing equation for $G^{(0)}$ and $G^{(1)}$ is also obtained.
The solution of $b^{(1)}$ is analytically obtained
\begin{align}
& b_i^{(1)}(\bm{p}) = 
 \int_{t_0}^t \mathrm{d}t^{\prime}
\sum_{\bm{p}^{\prime}}
G_{in}^{(0)}(\bm{p},t|\bm{p}^{\prime}, t^{\prime})
 \mathcal{N}_{njk}(\bm{p}_0,\bm{q}_0,\bm{r}_0)
\nonumber\\
&
\times \left(
\delta_{\bm{p}^{\prime} \bm{p}_0}
 b_j^{(0)}(-\bm{q}_0,t^{\prime}) b_k^{(0)}(-\bm{r}_0,t^{\prime})
+
\delta_{\bm{p}^{\prime} \bm{p}_0}
 b_j^{(0)}(\bm{q}_0,t^{\prime}) b_k^{(0)}(\bm{r}_0,t^{\prime})
\right)
\nonumber\\
&
\qquad\qquad
+
\{\bm{p}_0, \bm{q}_0, \bm{r}_0\}
\label{eq:b1}
\end{align}
under the initial condition $b_i^{(1)}(t=t^{\prime})=0$.
The solution of $G^{(1)}$ is also obtained analytically
and given by $b^{(0)}$ and $G^{(0)}$.

To investigate wave turbulent statistics,
the correlation between $b$'s
defined as
\begin{align*}
 \overline{V_{ij}(\bm{p}, t, t^{\prime})} &=
 \overline{b_i(\bm{p}, t) b_j^{\ast}(\bm{p}, t^{\prime})}
\end{align*}
is considered.
Because of Eq.~(\ref{eq:governingeqb})
the simultaneous correlation is governed by the following equation:
\begin{align}
 \frac{\partial \overline{V_{ij}(\bm{p}, t,t)}}{\partial t} &= \mathcal{L}_{im}(\bm{p}) \overline{V_{mj}(\bm{p},t,t)}
 +
\!\!\!\!\!\!\!\!
 \sum_{\bm{p} + \bm{q} + \bm{r} = \bm{0}}
\!\!\!\!\!\!
 \mathcal{N}_{imn}(\bm{p},\bm{q},\bm{r})
 \overline{b_j(-\bm{p}) b_m(-\bm{q}) b_n(-\bm{r})}
\nonumber\\
&
\qquad
+ \mathrm{c.c.}(i\leftrightarrow j)
.
\label{eq:autocorrelation0}
\end{align}
The summation of $m$ and $n$ is taken over $1$ and $2$.
DIA that removes the direct interactions among $\bm{p}+\bm{q}+\bm{r}=\bm{0}$
is applied to the wave field.
First,
Eq.~(\ref{eq:autocorrelation0}) is rewritten as
\begin{align}
& \frac{\partial \overline{V_{ij}(\bm{p}, t,t)}}{\partial t} = \mathcal{L}_{im}(\bm{p}) \overline{V_{mj}(\bm{p},t,t)}
\nonumber\\
&
 +
\!\!\!\!\!\!
 \sum_{\bm{p} + \bm{q} + \bm{r} = \bm{0}}
\!\!\!\!\!\!
 \mathcal{N}_{imn}(\bm{p},\bm{q},\bm{r})
\left(
\overline{b_j^{(0)}(-\bm{p}) b_m^{(0)}(-\bm{q}) b_n^{(0)}(-\bm{r})}
+ \overline{b_j^{(1)}(-\bm{p}) b_m^{(0)}(-\bm{q}) b_n^{(0)}(-\bm{r})}
\right.
\nonumber\\
& \left.
+ \overline{b_j^{(0)}(-\bm{p}) b_m^{(1)}(-\bm{q}) b_n^{(0)}(-\bm{r})}
+ \overline{b_j^{(0)}(-\bm{p}) b_m^{(0)}(-\bm{q}) b_n^{(1)}(-\bm{r})}
\right)
\nonumber\\
&
+ \mathrm{c.c.}(i\leftrightarrow j),
\label{eq:vtt_2nd}
\end{align}
by resolving $b$ in the right-hand side of Eq.~(\ref{eq:autocorrelation0})
into $b^{(0)}$ (NDI) and $b^{(1)}$ (DI).
In NDI field which has no interactions among $\bm{p}$, $\bm{q}$ and $\bm{r}$,
$\overline{b_j^{(0)}(-\bm{p}) b_m^{(0)}(-\bm{q}) b_n^{(0)}(-\bm{r})}=0$,
since
$b_j^{(0)}(\bm{p})$, $b_m^{(0)}(\bm{q})$, $b_n^{(0)}(\bm{r})$ are statistically independent.
Second,
the solutions~(\ref{eq:b1}) are substituted into
$\overline{b_j^{(1)}(-\bm{p}) b_m^{(0)}(-\bm{q}) b_n^{(0)}(-\bm{r})}$ and similar terms.
At last,
Eq.~(\ref{eq:autocorrelation0}) is expressed by
the different-time correlation and the response function:
\begin{align}
& \frac{\partial \overline{V_{ij}(\bm{p}, t,t)}}{\partial t} = \mathcal{L}_{im}(\bm{p}) \overline{V_{mj}(\bm{p},t,t)}
\nonumber\\
&
 +2
\!\!\!\!\!\!
 \sum_{\bm{p} + \bm{q} + \bm{r} = \bm{0}}
\!\!\!\!\!\!
 \mathcal{N}_{imn}(\bm{p},\bm{q},\bm{r})
\int_{t_0}^t \mathrm{d}t^{\prime}
\left(
 \mathcal{N}_{abc}(\bm{p},\bm{q},\bm{r})
\overline{ G_{ja}(-\bm{p},t | {-}\bm{p},t^{\prime}) }
\ 
\overline{ V_{bm}(\bm{q},t^{\prime},t) }
\ 
\overline{ V_{cn}(\bm{r},t^{\prime},t) }
\right.
\nonumber\\
&
+
 \mathcal{N}_{abc}(\bm{q},\bm{r},\bm{p})
\overline{ G_{ma}(-\bm{q},t | {-}\bm{q},t^{\prime}) }
\ 
\overline{ V_{bn}(\bm{r},t^{\prime},t) }
\ 
\overline{ V_{cj}(\bm{p},t^{\prime},t) }
\nonumber\\
&
\left.
+
 \mathcal{N}_{abc}(\bm{r},\bm{p},\bm{q})
\overline{ G_{na}(-\bm{r},t | {-}\bm{r},t^{\prime}) }
\ 
\overline{ V_{bj}(\bm{p},t^{\prime},t) }
\ 
\overline{ V_{cm}(\bm{q},t^{\prime},t) }
\right)
\nonumber\\
& + \mathrm{c.c.}(i\leftrightarrow j)
.
\label{eq:DIAsimul}
\end{align}
The statistical independence between $b_i$ and $G_{mn}$
is assumed.

A similar procedure can be applied to the different-time correlation.
Hence, the governing equation of the different-time correlation is obtained as
\begin{align}
& \frac{\partial \overline{V_{ij}(\bm{p}, t,t^{\prime})}}{\partial t} = \mathcal{L}_{im}(\bm{p}) \overline{V_{mj}(\bm{p},t,t^{\prime})}
\nonumber\\
&
 +2
\!\!\!\!\!\!
 \sum_{\bm{p} + \bm{q} + \bm{r} = \bm{0}}
\!\!\!\!\!\!
 \mathcal{N}_{imn}(\bm{p},\bm{q},\bm{r})
\left(
\int_{t_0}^{t^{\prime}} \mathrm{d}t^{\prime\prime}
 \mathcal{N}_{abc}(\bm{p},\bm{q},\bm{r})
\overline{ G_{ja}(-\bm{p},t^{\prime} | {-}\bm{p},t^{\prime\prime}) }
\ 
\overline{ V_{bm}(\bm{q},t^{\prime\prime},t) }
\ 
\overline{ V_{cn}(\bm{r},t^{\prime\prime},t) }
\right.
\nonumber\\
&
+
\int_{t_0}^t \mathrm{d}t^{\prime\prime}
\left(
 \mathcal{N}_{abc}(\bm{q},\bm{r},\bm{p})
\overline{ G_{ma}(-\bm{q},t | {-}\bm{q},t^{\prime\prime}) }
\ 
\overline{ V_{bn}(\bm{r},t^{\prime\prime},t) }
\ 
\overline{ V_{cj}(\bm{p},t^{\prime\prime},t^{\prime}) }
\right.
\nonumber\\
&
\left.
\left.
+
 \mathcal{N}_{abc}(\bm{r},\bm{p},\bm{q})
\overline{ G_{na}(-\bm{r},t | {-}\bm{r},t^{\prime\prime}) }
\ 
\overline{ V_{bj}(\bm{p},t^{\prime\prime},t^{\prime}) }
\ 
\overline{ V_{cm}(\bm{q},t^{\prime\prime},t) }
\right)
\right)
.
\label{eq:DIA2time}
\end{align}
Moreover,
the governing equation of the response function is also obtained as
\begin{align}
& \frac{\partial \overline{G_{in}(\bm{p}, t | \bm{p}, t^{\prime})}}{\partial t}
 = \mathcal{L}_{ij}(\bm{p}) \overline{G_{jn}(\bm{p}, t | \bm{p}, t^{\prime})}
\nonumber\\
 &+ 4
 \!\!\!\!\!\!
 \sum_{\bm{p} + \bm{q} + \bm{r} = \bm{0}}
 \!\!\!\!\!\!
 \mathcal{N}_{ijk}(\bm{p},\bm{q},\bm{r})
 \mathcal{N}_{abc}(\bm{r},\bm{p},\bm{q})
\int_{t^{\prime}}^{t} \mathrm{d}t^{\prime\prime}
 \overline{G_{ka}(-\bm{r}, t | {-}\bm{r}, t^{\prime\prime})}
\ 
 \overline{G_{bn}(\bm{p}, t^{\prime\prime} | \bm{p}, t^{\prime})}
\ 
\overline{V_{cj}(\bm{q}, t^{\prime\prime}, t)}
.
\label{eq:DIAgreen}
\end{align}
Equations~(\ref{eq:DIAsimul}--\ref{eq:DIAgreen})
give a closed equation system
and they are the DIA equations of the wave turbulence.

It should be emphasized
that the weak nonlinearity is not assumed in the procedure.
Therefore,
the DIA equations of the wave turbulence (\ref{eq:DIAsimul}--\ref{eq:DIAgreen}) can be applied
also to strongly nonlinear wave turbulent systems.

\section{DIA Equations for Autocorrelation of Complex Amplitude}

In the previous section,
DIA is applied to the variable $b$.
Turbulent statistics is described with the complex amplitude $a$
to compare the weak turbulence theory
in this section.
The correlation of $a$ is defined as
$\overline{a(\bm{p}, t) a^{\ast}(\bm{p}, t^{\prime})} = N(\bm{p}, t, t^{\prime})$
and
$\overline{a(\bm{p}, t) a(-\bm{p}, t^{\prime})} = M(\bm{p}, t, t^{\prime})$.
The correlation of $b$, that is $\overline{V_{ij}}$
is expressed by $M$ and $N$
as
 \begin{align*}
  & \overline{V_{ij}(\bm{p}, t, t^{\prime})} =
  \frac{1}{4} \left(
  \mathrm{i}^{-(i-j)} N(\bm{p}, t, t^{\prime})
  +
  \mathrm{i}^{i-j}
  N^{\ast}(-\bm{p}, t, t^{\prime})
\right.
\nonumber\\
&
\qquad\qquad\qquad
\left.
  - \left(
  \mathrm{i}^{-(i+j)} M(\bm{p}, t, t^{\prime})
  +
  \mathrm{i}^{i+j} M^{\ast}(-\bm{p}, t, t^{\prime})
  \right)
  \right)
.
\end{align*}
The initial condition of the cross-correlation is that $M(\bm{p}, t_0, t_0) = 0$
since $a(\bm{p})$ and $a(-\bm{p})$ are uncorrelated initially.
The cross-correlation at later time
is much smaller than the auto-correlation,
$|M(\bm{p}, t, t^{\prime})| \ll |N(\bm{p}, t, t^{\prime})|$.
Therefore,
\begin{align*}
 & \overline{V_{ij}(\bm{p}, t| \bm{p}, t^{\prime})} =
 \frac{1}{4} \left(
 \mathrm{i}^{-(i-j)} N(\bm{p}, t, t^{\prime})
 +
 \mathrm{i}^{i-j}
 N^{\ast}(-\bm{p}, t, t^{\prime})
 \right)
,
\nonumber\\
 & \overline{G_{ij}(\bm{p}, t| \bm{p}, t^{\prime})} =
 \frac{1}{2} \left(
 \mathrm{i}^{-(i-j)} G(\bm{p}, t, t^{\prime})
 +
 \mathrm{i}^{i-j}
 G^{\ast}(-\bm{p}, t, t^{\prime})
 \right)
.
\end{align*}

Equation~(\ref{eq:DIAsimul}) is rewritten to a equation for $N(\bm{p}, t, t)$
as
\begin{align}
& \frac{\partial N(\bm{p}, t, t)}{\partial t} =
2
\!\!\!\!\!\!\!\!
\sum_{\bm{p} + \bm{q} + \bm{r} = \bm{0}}
\int_{t_0}^{t} \mathrm{d}t^{\prime}
\left(
G^{\ast}(\bm{p}, t, t^{\prime})
\right.
\left.
\left(
|\mathcal{T}^{\bm{p}}_{-\bm{q}-\bm{r}} |^2 N^{\ast}(-\bm{q}, t^{\prime}, t) 
N^{\ast}(-\bm{r}, t^{\prime}, t)
\right.
\right.
\nonumber\\
&
\left.
+
|\mathcal{T}^{-\bm{q}}_{\bm{r}\bm{p}} |^2 N^{\ast}(-\bm{q}, t^{\prime}, t) N(\bm{r}, t^{\prime}, t)
+
|\mathcal{T}^{-\bm{r}}_{\bm{p}\bm{q}} |^2 N(\bm{q}, t^{\prime}, t) N^{\ast}(-\bm{r}, t^{\prime}, t)
\right)
\nonumber\\
&
-
N(\bm{p}, t^{\prime}, t)
\left(
|\mathcal{T}^{\bm{p}}_{-\bm{q}-\bm{r}} |^2 
G(-\bm{q}, t, t^{\prime})
N^{\ast}(-\bm{r}, t^{\prime}, t) 
\right.
\nonumber\\
&\left.
+
|\mathcal{T}^{-\bm{q}}_{\bm{r}\bm{p}} |^2 
G(-\bm{q}, t, t^{\prime})
N(\bm{r}, t^{\prime}, t) 
-
|\mathcal{T}^{-\bm{r}}_{\bm{p}\bm{q}} |^2 
G^{\ast}(\bm{q}, t, t^{\prime})
N^{\ast}(-\bm{r}, t^{\prime}, t) 
\right)
\nonumber\\
&
\left.
+ \{\bm{q} \leftrightarrow \bm{r}\}
\right)
+ \mathrm{c.c.}
\label{eq:DIAnsimul}
\end{align}
Similarly,
the different-time correlation $N(\bm{p}, t, t^{\prime})$
is also obtained from Eq.~(\ref{eq:DIA2time}).
Moreover,
Eq.~(\ref{eq:DIAgreen}) can be expressed by $N$ and $G$.
Namely,
the DIA equations~(\ref{eq:DIAsimul}--\ref{eq:DIAgreen})
in the wave turbulence
as equations for the correlation $\overline{V_{ij}}$ of $b$
can be rewritten as equations for the correlation $N$ for $a$.

Furthermore,
the fluctuation--dissipation relation,
$N(\bm{p}, t, t^{\prime}) = n(\bm{p}, t^{\prime}) G(\bm{p}, t, t^{\prime})$,
is employed.
The simultaneous correlation is written as $N(\bm{p}, t, t) = n(\bm{p}, t)$ from now on.
Hence,
the DIA equations~(\ref{eq:DIAsimul}--\ref{eq:DIAgreen})
can be rewritten as
\begin{align}
& \frac{\partial n(\bm{p}, t)}{\partial t} =
2
\!\!\!\!\!\!\!\!
\sum_{\bm{p} + \bm{q} + \bm{r} = \bm{0}}
\int_{t_0}^{t} \mathrm{d}t^{\prime}
\left(
|\mathcal{T}^{\bm{p}}_{-\bm{q}-\bm{r}} |^2
\left(
n(-\bm{q}, t^{\prime}) 
n(-\bm{r}, t^{\prime})
- n(\bm{p}, t^{\prime}) 
\left(
n(-\bm{q}, t^{\prime}) 
+
n(-\bm{r}, t^{\prime})
\right)
\right)
\right.
\nonumber\\
&
\left.
G^{\ast}(\bm{p}, t, t^{\prime})
G(-\bm{q}, t, t^{\prime}) 
G(-\bm{r}, t, t^{\prime})
\right.
\nonumber\\
&
\left.
-
|\mathcal{T}^{-\bm{q}}_{\bm{r}\bm{p}} |^2
\left(
 n(\bm{p}, t^{\prime}) n(\bm{r}, t^{\prime})
-
 n(-\bm{q}, t^{\prime})
\left(
 n(\bm{p}, t^{\prime}) + n(\bm{r}, t^{\prime})
\right)
\right)
\right.
\left.
 G^{\ast}(\bm{p}, t, t^{\prime})
 G(-\bm{q}, t, t^{\prime})
 G^{\ast}(\bm{r}, t, t^{\prime})
\right.
\nonumber\\
&
\left.
-
|\mathcal{T}^{-\bm{r}}_{\bm{p}\bm{q}} |^2
\left(
 n(\bm{p}, t^{\prime}) n(\bm{q}, t^{\prime})
-
 n(-\bm{r}, t^{\prime})
\left(
 n(\bm{p}, t^{\prime}) + n(\bm{q}, t^{\prime})
\right)
\right)
\right.
\left.
 G^{\ast}(\bm{p}, t, t^{\prime})
 G^{\ast}(\bm{q}, t, t^{\prime})
 G(-\bm{r}, t, t^{\prime})
\right)
\nonumber\\
&
+ \mathrm{c.c.}
\label{eq:DIAnsimulfd}
\end{align}
and
\begin{align}
& \frac{\partial G(\bm{p}, t, t^{\prime})}{\partial t} =
- \mathrm{i} \omega(\bm{p}) G(\bm{p}, t, t^{\prime})
\nonumber\\
&
- 2
\!\!\!\!\!\!\!\!
\sum_{\bm{p} + \bm{q} + \bm{r} = \bm{0}}
\int_{t^{\prime}}^{t} \mathrm{d}t^{\prime\prime}
G(\bm{p}, t^{\prime\prime}, t^{\prime})
\left(
|\mathcal{T}^{\bm{p}}_{-\bm{q} -\bm{r}} |^2 
\left(
n(-\bm{q}, t^{\prime\prime})
+
n(-\bm{r}, t^{\prime\prime})
\right)
G(-\bm{q}, t, t^{\prime\prime}) 
G(-\bm{r}, t, t^{\prime\prime})
\right.
\nonumber\\
&
\qquad\qquad\qquad
-
|\mathcal{T}^{-\bm{q}}_{\bm{r}\bm{p}} |^2 
\left(
n(-\bm{q}, t^{\prime\prime}) 
- n(\bm{r}, t^{\prime\prime}) 
\right)
G(-\bm{q}, t, t^{\prime\prime})
G^{\ast}(\bm{r}, t, t^{\prime\prime})
\nonumber\\
&
\left.
\qquad\qquad\qquad
-
|\mathcal{T}^{-\bm{r}}_{\bm{p}\bm{q}} |^2 
\left(
n(-\bm{r}, t^{\prime\prime}) 
- n(\bm{q}, t^{\prime\prime}) 
\right)
G^{\ast}(\bm{q}, t, t^{\prime\prime})
G(-\bm{r}, t, t^{\prime\prime})
\right)
.
\label{eq:DIAg}
\end{align}
Here, $t_0 = t^{\prime}$ thanks to the causality.
Equations~(\ref{eq:DIAnsimulfd}) and (\ref{eq:DIAg}) are another set of the DIA equations.

Equation~(\ref{eq:DIAnsimulfd}) in the DIA equations and the conventional kinetic equation
look quite similar.
Nonetheless, there exist some differences.
Hysteresis is included in the DIA equations 
since the time integration is incorporated in Eq.~(\ref{eq:DIAnsimulfd}).
It is in contrast with the Markovian properties of the kinetic equation.
Furthermore,
the response functions appear in the integrand of Eq.~(\ref{eq:DIAnsimulfd}).
Hence,
the DIA equations accept nonlinear changes of phases.
The coherent structures such as freak waves which develop abruptly
have the non-Markovian properties and the phase entrainment.
Therefore, it is expected that
the intermittency is evaluated statistically by the DIA equations.

The quadratic energy $\sum_{\bm{p}} \omega(\bm{p}) n(\bm{p})$
is not strictly conserved for Eq.~(\ref{eq:DIAnsimulfd})
since Hamiltonian which is the sum of quadratic and cubic energies is conserved.
The quadratic energy is conserved for the kinetic equation~(\ref{eq:kineticeq})
owing to the resonant interactions.
Therefore,
this non-conservation laws implies that
the fast non-resonant interactions associated with strongly nonlinear coherent structures
can be statistically estimated by the nonlinear parts of the response functions.
The numerical simulations of Eqs.~(\ref{eq:DIAnsimulfd}) and (\ref{eq:DIAg})
are required to evaluate the fast non-resonant interactions
for the specific wave turbulent systems.
The quadratic momentum $\sum_{\bm{p}} \bm{p} n(\bm{p})$
is a conserved quantity
because of the conditions of the wavenumbers.
Therefore,
one can find an equilibrium solution $n(\bm{p}) \propto (\bm{p} \cdot \bm{U})^{-1}$,
where $\bm{U}$ is an arbitrary constant vector.
The entropy $\sum_{\bm{p}} \log n(\bm{p})$ never decrease in Eq.~(\ref{eq:DIAnsimulfd}).
In this manner,
some of statistical natures
are the same as the conventional kinetic equation (\ref{eq:kineticeq}).

For the short-time limit $t \to t_0$,
Eq.~(\ref{eq:DIAnsimulfd})
is rewritten as
\begin{align}
\frac{\partial n(\bm{p})}{\partial t} \approx &
4 (t - t_0)
\!
\left(
\!
 \sum_{\bm{p} = \bm{q} + \bm{r}}
\!\!\!\!\!
\left|\mathcal{T}^{\bm{p}}_{\bm{q} \bm{r}}\right|^2
\left(n(\bm{q}) n(\bm{r}) - n(\bm{p}) (n(\bm{q}) + n(\bm{r}))\right)
\right.
\nonumber\\
&
\quad
-
\!\!\!\!
 \sum_{\bm{q} = \bm{r} + \bm{p}}
\!\!\!\!
\left|\mathcal{T}^{\bm{q}}_{\bm{r} \bm{p}}\right|^2
\left(n(\bm{r}) n(\bm{p}) - n(\bm{q}) (n(\bm{r}) + n(\bm{p}))\right)
\nonumber\\
&
\quad
\left.
-
\!\!\!\!
 \sum_{\bm{r} = \bm{p} + \bm{q}}
\!\!\!\!
\left|\mathcal{T}^{\bm{r}}_{\bm{p} \bm{q}}\right|^2
\left(n(\bm{p}) n(\bm{q}) - n(\bm{r}) (n(\bm{p}) + n(\bm{q}))\right)
\right)
.
\label{eq:shorttime}
\end{align}
This is also consistent with the short-time kinetic equation in Ref.~\cite{janssen_freak}.
Equation~(\ref{eq:shorttime}) is valid even for strongly nonlinear regimes
since the equation is derived
without employing the separation of the linear and nonlinear time scales.

\section{Recovery of Kinetic Equation from DIA Equations}
To recover the kinetic equation from Eq.~(\ref{eq:DIAnsimul}),
the weak nonlinearity is assumed as an ``extra'' assumption in this section.
The time variations of the different-time correlation and the response function
are as follows:
\begin{align*}
& \frac{\partial N(\bm{p}, t, t^{\prime})}{\partial t} = -\mathrm{i} \omega N(\bm{p}, t, t^{\prime})
+ \mathrm{nonlinear\ terms}
,
\\
& \frac{\partial G(\bm{p}, t, t^{\prime})}{\partial t} = -\mathrm{i} \omega G(\bm{p}, t, t^{\prime})
+ \mathrm{nonlinear\ terms}
.
\end{align*}
The nonlinear terms of the different-time correlation and the response function
do not contribute to the simultaneous correlation at the leading order.
Then, by neglecting the nonlinear terms,
the leading terms of the different-time correlation and the response function
are obtained as
 \begin{align*}
& N(\bm{p}, t, t^{\prime}) = n(\bm{p}, t^{\prime})
 \mathrm{e}^{-\mathrm{i} \omega (t-t^{\prime})}
,
\quad
G(\bm{p}, t, t^{\prime}) =
\mathrm{e}^{-\mathrm{i} \omega (t-t^{\prime})}
,
\end{align*}
under the assumption of the weak nonlinearity.
Therefore,
since $n(\bm{p}, t)$ varies much slower
than $1/\Delta \omega$,
$n(\bm{p}, t) = n(\bm{p}, t^{\prime}) = n(\bm{p})$.
This $n(\bm{p})$ is the very wave action in the weak turbulence theory.
By substituting the leading terms of the different-time correlation and the response function into Eq.~(\ref{eq:DIAnsimul}),
we obtain
\begin{align}
& \frac{\partial n(\bm{p})}{\partial t} \! = \!
2
\!\!
\int_{t_0}^t \! \mathrm{d}t^{\prime}
\!
\left(
\!
 \sum_{\bm{p} = \bm{q} + \bm{r}}
\!\!\!\!\!
\left|\mathcal{T}^{\bm{p}}_{\bm{q} \bm{r}}\right|^2
\left(n(\bm{q}) n(\bm{r}) - n(\bm{p}) (n(\bm{q}) + n(\bm{r}))\right)
\right.
\left.
\mathrm{e}^{\mathrm{i} \Delta \omega^{\bm{p}}_{\bm{q} \bm{r}} (t-t^{\prime})}
\right.
\nonumber\\
&
-
\!\!\!\!
 \sum_{\bm{q} = \bm{r} + \bm{p}}
\!\!\!\!
\left|\mathcal{T}^{\bm{q}}_{\bm{r} \bm{p}}\right|^2
\left(n(\bm{r}) n(\bm{p}) - n(\bm{q}) (n(\bm{r}) + n(\bm{p}))\right)
\mathrm{e}^{-\mathrm{i} \Delta \omega^{\bm{q}}_{\bm{r} \bm{p}} (t-t^{\prime})}
\nonumber\\
&
\left.
-
\!\!\!\!
 \sum_{\bm{r} = \bm{p} + \bm{q}}
\!\!\!\!
\left|\mathcal{T}^{\bm{r}}_{\bm{p} \bm{q}}\right|^2
\left(n(\bm{p}) n(\bm{q}) - n(\bm{r}) (n(\bm{p}) + n(\bm{q}))\right)
\mathrm{e}^{-\mathrm{i} \Delta \omega^{\bm{r}}_{\bm{p} \bm{q}} (t-t^{\prime})}
\right)
\nonumber\\
& + \mathrm{c.c.}
\label{eq:kinetic-1}
\end{align}
Since the separation of the time scales is assumed,
$t_0$ can be set to $-\infty$.
Then,
\begin{align}
\int_{t_0}^{t} \mathrm{d}t^{\prime } \mathrm{e}^{\mathrm{i} \Delta \omega^{\bm{p}}_{\bm{q}\bm{r}} (t-t^{\prime})}
= \mathrm{i} (\mathrm{P.V.}\left(\frac{1}{\Delta \omega^{\bm{p}}_{\bm{q}\bm{r}}}\right)
 - \mathrm{i} \pi \delta(\Delta \omega^{\bm{p}}_{\bm{q}\bm{r}}) ).
\label{eq:convergencedelta}
\end{align}
Finally,
Eq.~(\ref{eq:kinetic-1}) results in Eq.~(\ref{eq:kineticeq}).
Namely,
the kinetic equation in the weak turbulence theory can be recovered 
by an additional assumption, that is, the weak nonlinearity to the DIA equations.

\section{Concluding Remark}
\label{sec:conclusion}

\subsection{Discussion}
\label{sec:discussion}

In the short-time limit,
Eq.~(\ref{eq:shorttime}) is consistent with
what is derived along the weak turbulence theory.
The conventional kinetic equation is also recovered
from the DIA equations in the weakly nonlinear limit.
The nonlinear parts of the response function
are irrelevant to the time variation of the wave action in both limits.

As pointed out in direct numerical simulations of four-wave weak turbulent system~\cite{doi:10.1175/JPO3029.1},
the convergence to the kinetic equation is much faster
than the convergence of the integral to the $\delta$ function in Eq.~(\ref{eq:convergencedelta}).
During the intermediate time,
which is longer than the short time and shorter than the convergence,
the nonlinear parts of the response function will play a role.
The role will be evaluated for specific systems with numerical simulations.

\subsection{Conclusion}
\label{subsec:conclusion}

The closed equation system is developed for the not-necessarily-weak wave turbulence statistics
according to direct-interaction approximation (DIA).
In the procedure,
the three assumptions below are made:
\begin{itemize}
 \item Quantities in the field without the direct interactions are much larger than that in the perturbed field
under the largeness of the degrees of freedom in the wave field.
 \item $b_j^{(0)}(\bm{p})$, $b_m^{(0)}(\bm{q})$ and $b_n^{(0)}(\bm{r})$ are statistically independent in the field without the direct interactions among
       $\bm{p}$, $\bm{q}$ and $\bm{r}$.
 \item $b_i$ and $G$ are statistically independent.
\end{itemize}
By developing the equation system without assuming the weak nonlinearity,
the DIA equations can consistently make
statistical description for the wave turbulent system
that has even short-time and strong nonlinear interactions.
It provides us an appropriate tool to evaluate the intermittency in the wave turbulent system.

The kinetic equation in the weak turbulence theory is also recovered
from the DIA equations.
This indicates that
the framework of DIA which harness the largeness of the degrees of freedom
is the natural extension of the weak turbulence theory.

\section*{Acknowledgments}
We gratefully thank Dr.~Goto for valuable discussions on the direct-interaction approximation.

\end{document}